# Microstructure of erosion spots on the surface interacting with filamented beam plasma

V.A. Rantsev-Kartinov

NFI RRC "Kurchatov Institute", Moscow 123182, Russia

The fine structure of micron-sized erosion spots within coaxial rings of 10-100 micron diameter, observed in the Mather-type plasma focus, is analyzed. The topological similarity of this structuring to that of electric current filaments, observed in the Filippov-type plasma foci and straight Z-pinches, and dust particles deposits in tokamak T-10, is shown. The possibility of interpreting this structuring in terms of the formerly suggested hypothesis for long-lived electric current filament formation, due to electrodynamic aggregation of nanodust in electric discharges, is discussed.

## 1. INTRODUCTION

A detailed analysis of micro- and nanostructure of various carbonaceous and hydrocarbonaceous films and agglomerates, deposited on the surface of vacuum chamber and special collectors inside the chamber in tokamak T-10, carried out in [1-4], revealed the identity of its topological structuring to that of the following structures: coaxial tubular blocks, sometimes with radial spokes, and cartwheel-like structures as a separate block on its own axle or that in the but-end of the tubule. That analysis has continued
- the revelation of long-lived filamentary structures in the images of the plasma in peripheral plasma and edge plasma in various thermonuclear fusion devices, and
- the formulation of the hypothesis for an electrodynamic aggregation of nanodust (first of all, carbon nanotubes or similar nanostructures composed of atoms of other chemical elements) to give long-lived macroscopic fractal structures (with the basic block of tubular form) which serve as a skeleton of the observed filamentary structures, at least in peripheral plasmas (see short summary of the phenomenon in [5], the surveys [6-8] and web page [9]).

The processes of erosion of the vacuum chamber's surface and plasma limiters' surface, and the material's deposition on these surfaces, may also be influenced by this phenomenon. First, the eroded surface or the deposited structures on the surface may be an imprint of the filamentary structuring in electric current-carrying plasmas. And second, the beam-surface interaction may produce the conditions for electrodynamic aggregation of macroscopic structures of the above mentioned topology on the surface. Here we analyze these points on the example of data from the Mather-type plasma focus facility.

## 2. The electron beam-produced structures on the surface

The fine structuring produced by the electron beam, incident on the surface of the anode in the Mather-type plasma focus facility and escaped from the chamber through the hole to interact with a target, has been found and analyzed in [10]. The observed structure was a set of coaxial rings composed of isolated, almost identical spots (jags) – see Figs. 1a and 2a. Such a structuring of the eroded surface of the anode was interpreted as a result of microbursts on the surface under the action of an electron beam, i.e. a sort of the "etching" of the surface by an electron beam.

The filamentary structure of plasma luminosity of the electric current sheath in plasma focus facilities is well recognized. The fine structure of electric current in a single filament of luminosity may coincide with the fine structure of electron beams escaped from the chamber, whatever would be the mechanisms of beam generation within the filament (e.g. via conversion of magnetic energy into electron acceleration by a local electric field). In particular, the filamentary network may have a complicated structure like, e.g., that of a closed magnetic configuration, namely, a filamentary spheromak-like structure, formed presumably via the reconnection of electric current (and magnetic field) within the entire electric current sheath and subsequently compressed by the magnetic field of the residual electric current [11-14]. The imprinting of the structure of electric current filaments seems to be of universal nature, visible both on the anode's surface (if the filament closes the electric circuit) and on the other parts of vacuum chamber's wall (e.g., if the straight section of the decaying filament generates an electron beam).

Here we show the results of processing (with the help of the method of multilevel dynamical contrasting [15, 16]) of the images [10] of the surface interacted with the electron beam produced in the Mather-type plasma focus facility in electric discharges with hydrogen working gas. The eroded surface is that of a plastic materials (with a high value of the sublimation, ~ 90 kcal/g atom) located behind an axial aperture in the anode on the distance of 6-300 cm from its surface, i.e. from the beam source (see [10] for more detail). The results of processing of these images at different distances from the anode are shown in Figures 1 and 2.

The individual filaments have produced the erosion spots located rather stochastically (see Fig. 2 in [10]). The imprint of a single filament appears to have a distinct hollow structure that may correspond to a tubular structure of the electric current within a single filament. Moreover, the substructure of such a filament appears to be a set of coaxial (nested) rings (Fig. 2) that may correspond to a distinct nested structure of the current within the filament. On the whole, this structuring coincides with that (namely, tubularity, sometimes coaxial one) revealed in a wide range of data from electric discharge facilities (see [6-9]).

The fine structure of the interaction of the single filament with the surface appears to produce the conditions for electrodynamic aggregation of the macroscopic structures of the above mentioned topology. Indeed, one can see coaxial tubular blocks, sometimes with radial spokes, and cartwheel-like structures as a separate block on its own axle or that in the but-end of the tubule.

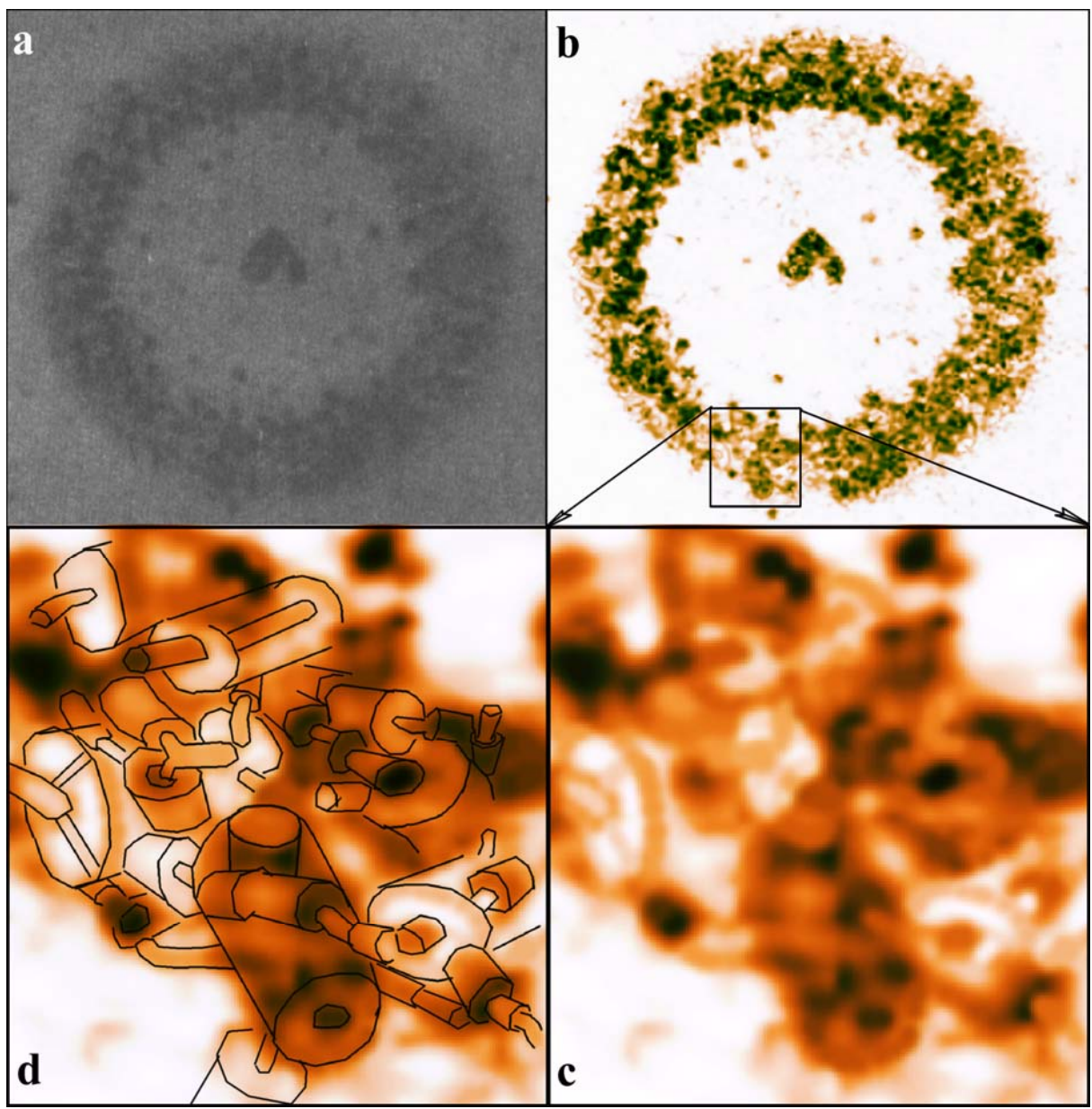

Fig. 1. The results of processing the image of Fig. 3a in [10] with the help of the method of multilevel dynamical contrasting [15, 16]. The width of the original image (figure "a") is ~ 40 microns. Figure "b" shows the MDC-processed image, the window (figure "c") is ~ 7 microns wide. The figure "d" gives a schematic drawing of the structure seen in image of figure "c" (reconstructed with the help of mosaic MDC method, see [3]). The tubular blocks seen in figure "c" are of ~ 1 micron diameter.

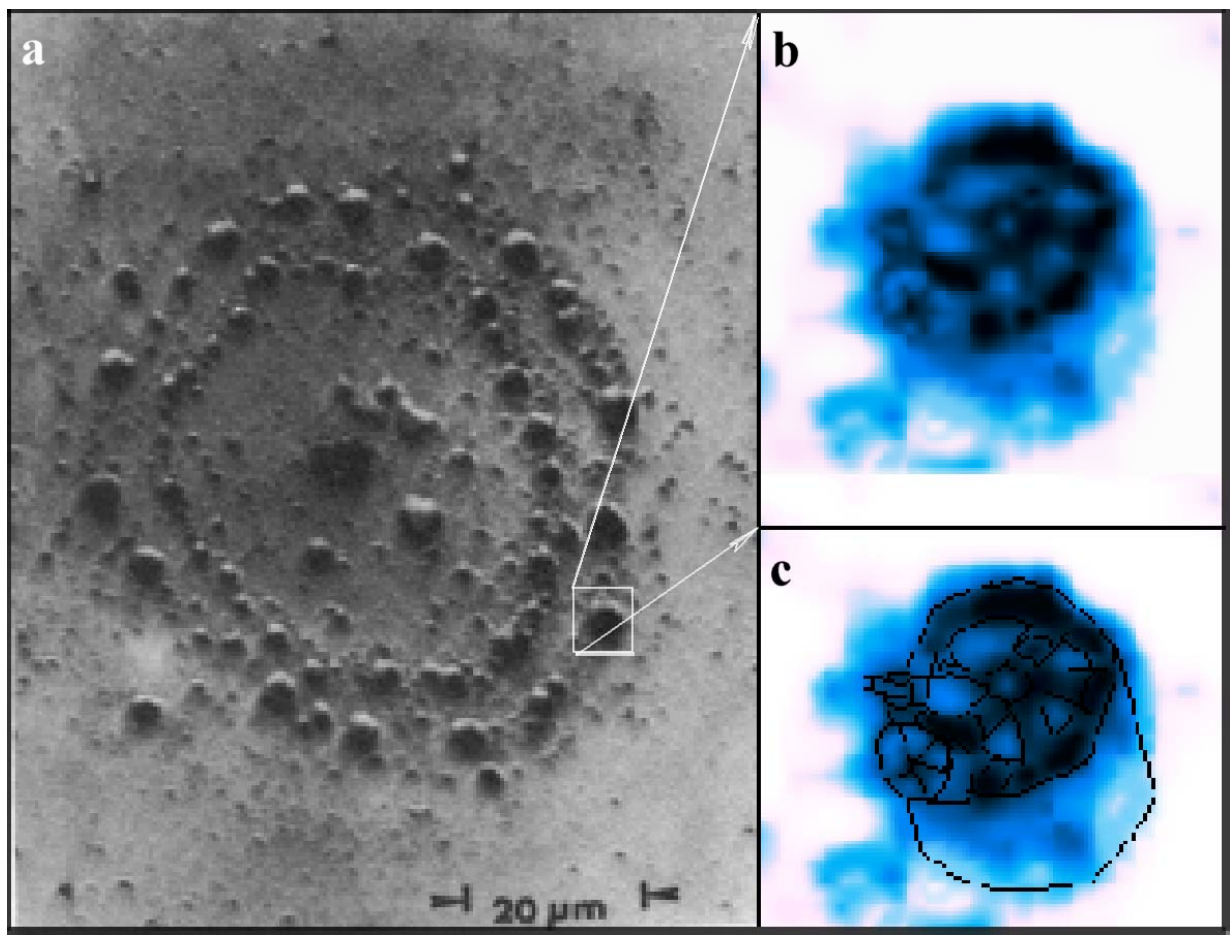

Fig. 2. (a) Original image, Fig. 3b in [10], of the eroded surface, seen as a coaxial rings of blobs, at first glance, spherical ones.(b) The MDC-processed image in the window, which shows the presence of a coaxial tubular skeletal structure with a cartwheel in the butt-end of diameter ~ 4 microns. (c) Schematic drawing of the figure "b" obtained with the help of mosaic MDC method.

## 3. Conclusions

An analysis of the fine structure of the erosion spots on the surface of the target, irradiated with the electron beam from the Mather-type plasma focus facility, shows the topological identity of the revealed structuring with that of the electric current filaments observed in the Filippov-type plasma foci and straight Z-pinches, and that of dust particles deposits in tokamak T-10.

The interpretation of this structuring in terms of the formerly suggested hypothesis for long-lived electric current filament formation, due to electrodynamic aggregation of nanodust in electric discharges, may be based on the key role of electrodynamic processes in an expanding plasma produced by the irradiation of a condensed matter target by a pulsed electron beam (cf. similar interpretation [6] for the case of irradiating a target with a laser pulsed beam). Indeed, the expanding plasma may produce an electric current directed opposite to electron beam. This pulsed current produces a strong magnetic field which, in turn, may produce a return electric current and thus close an effective electric circuit at the periphery of the plasma corona. The central near-axis region of the expanding plasma corona possesses the properties of a Z-pinch or plasma focus plasma column and may produce skeletal fractal structures with the basic block of tubular form.

**Acknowledgments**

This work is supported by the Russian Foundation for Basic Research (project RFBR 05-08-65507).